# Half-percent terahertz generation efficiency from cryogenically cooled lithium niobate pumped by Ti:sapphire laser pulses

Xiaojun Wu,[1,3,*] Koustuban Ravi,[1,4] Wenqian Ronny Huang,[4] Chun Zhou,[1,2] Peter Zalden,[3,5] Giulio M. Rossi,[1,3] Giovanni Cirmi,[1,3] Oliver D. Mücke,[1,3] and Franz X. Kärtner[1,2,3,4]

[1]Center for Free-Electron Laser Science, DESY, Notkestraße 85, Hamburg 22607, Germany
[2]Department of Physics, University of Hamburg, Hamburg 22761, Germany
[3]The Hamburg Center for Ultrafast Imaging, Luruper Chaussee 149, Hamburg 22761, Germany
[4]Department of Electrical Engineering and Computer Science, and Research Laboratory of Electronics, Massachusetts Institute of Technology, 77 Massachusetts Avenue, Cambridge, Massachusetts 02139, USA
[5]European XFEL, Albert-Einstein-Ring 19, Hamburg 22761, Germany
[*]Corresponding author: xiaojun.wu@desy.de

**We obtained an optical-to-terahertz (THz) energy conversion efficiency of 0.5% using the tilted-pulse-front technique in lithium niobate at a cryogenically cooled temperature of 100 K pumped by amplified Ti:sapphire laser pulses with ~150 fs pulse duration at 800 nm wavelength. Compared with the optimized conversion efficiency of 0.18% achieved at room temperature, we achieved more than 2.5 times enhancement in conversion efficiency upon cryogenically cooling the crystal due to reduction of THz absorption. Further improvements to the conversion efficiency can be made by optimizing the out-coupling of the THz radiation, transportation of pump energy and by further decreasing the THz absorption in the lithium niobate crystal.**

http://dx.doi.org/

Strong-field terahertz (THz) sources have experienced rapid development during recent years. Large facilities, such as free-electron lasers and laser-driven ion accelerators can generate up to 600 µJ [1] and 460 µJ [2] of THz pulse energy respectively. However, these THz sources are expensive and relatively challenging to control. Conventional table-top THz sources based on low-temperature grown GaAs antennae using the photoconductive effect can generate up to 1 µJ of THz output energy with 1.6% optical-to-THz energy conversion efficiency (henceforth referred to as conversion efficiency) [3]. However, the THz energy from a single device is limited as neither higher electric fields nor larger pump powers can be afforded due to breakdown and free-carrier screening effect limitations. Optical rectification (OR) in either organic or lithium niobate crystals is the most promising method for generation of extremely strong-field THz pulses. Although a conversion efficiency of 3.0% and THz pulse energies of 0.9 mJ has been demonstrated with organic crystals [4], this approach is limited by low damage thresholds and small crystal sizes. Furthermore, less accessible pump wavelengths ranging between 1.2-1.5 µm are required to pump these materials. Based on these challenges, the tilted-pulse-front technique with lithium niobate crystals still holds the most promise for generating mJ-level or even higher energy THz pulses [5]. Within this category, THz generation using 800 nm pump pulses is of particular interest due to the maturity of Ti:sapphire based laser technology.

Building on our previous work [6], we further optimized the tilted-pulse-front setup using a lithium niobate crystal pumped by Ti:sapphire laser pulses. At room temperature, we improved the conversion efficiency from 0.1% to 0.18% and further enhanced it to 0.5% by cryogenically cooling the crystal. To the best of our knowledge, this is the highest conversion efficiency from lithium niobate crystals pumped by Ti:sapphire laser pulses.

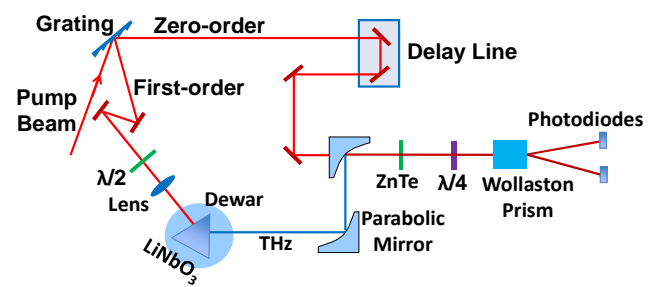

Fig. 1. Schematic of the tilted-pulse-front THz generation setup exploiting cryogenic cooling of the lithium niobate crystal.

In our experiment, we used a commercial cryogenic single-pass Ti:sapphire amplifier (Coherent Legend Elite Cryo OPA) producing pump pulses with a transform limited (TL) full-width at half-maximum (FWHM) pulse duration of ~150 fs at 1 kHz repetition rate. In Fig.

1, the pump beam entering the setup was collimated with a 13 mm (1/e$^2$ Gaussian) diameter. The pulse front was tilted using the first diffractive order of a gold-coated blazed grating with a groove density of 2000 lines/mm and diffraction efficiency of 85% at minus one order. The incident angle was 61.0° and the diffraction angle was 46.5°. One spherical lens of focal length 150 mm was used to image the titled-pulse-front onto the lithium niobate crystal. The polarization direction of the diffracted beam was rotated from the horizontal to vertical direction in order to align it with the optical axis of the crystal and achieve maximum second order nonlinearity. The lens to grating distance was 450 mm and crystal to lens distance was 225 mm. The demagnification factor is therefore -0.5. The crystal was a z-cut prism with base triangle dimensions of 13.8×13.8×13.0 mm, height of 12.7 mm and apex angle of 56.0°. The crystal was doped with 5.0% MgO in order to increase the damage threshold and reduce photorefractive effects. The crystal was cooled from room temperature down to 100 K as measured on the crystal surface with a calibrated silicon sensor (Lakeshore DT471SD) in a custom designed cryogenic chamber. The infrared (IR) input windows of the chamber were made of anti-reflection coated fused silica and the THz was extracted from a TPX (ploymethylpentene) window of 5 mm thickness and 88% transmissivity. The THz output facet of the prism is anti-reflection coated for the THz wave with a polyimide layer of 125 μm thickness which resulted in a 40% enhancement of the out-coupled THz energy for all the measurements. The output THz energy was measured with two bi-convex TPX lenses (50 mm focal length, 77% transmission) using a calibrated fast pyroelectric detector (Microtech Instruments) and compared against the results from a THz power meter (Ophir-Spiricon Model 3A-P-THz). Both detectors were covered by black plastic with 50% THz power transmission to block the scattered light from the pump beam. We employed a pyroelectric camera (Spiricon Pyrocam IV) covered by a 2-mm-thick silicon wafer to measure the imaged THz beams. The zero order reflected beam from the diffraction grating was used as probe for electro-optic sampling in a 0.2-mm-thick, (110)-oriented ZnTe crystal.

At room temperature, we optimized the THz generation setup at a peak pump fluence of 32 mJ/cm$^2$ corresponding to 7.1 mJ of pump energy on the crystal. The corresponding peak intensity of the pump is 214 GW/cm$^2$. Here, optimization of the setup is performed by careful adjustment of the tilt-angle by varying the lens to grating and lens to crystal distances, interaction length by varying the beam position with respect to the crystal edge as well as collection of the generated terahertz radiation. THz output energy of 13 μJ with corresponding conversion efficiency of 0.18% at 7.1 mJ pump energy was obtained. Fig. 2(a) shows the dependence of the extracted THz output energy and corresponding efficiency on the pump energy. Note that the optimized condition is slightly different for different pump energies and the above curve was acquired by maximizing the conversion efficiency for an input energy of 7.1 mJ and then decreasing the pump energy. We did not optimize the setup for each measurement point. The extracted THz energy exhibits a quadratic behavior with respect to the pump energy up to ~4.0 mJ (pump fluence: 18 mJ/cm$^2$) and then increases approximately linearly with pump energy, while the corresponding efficiency first increases linearly but starts to saturate at ~4.0 mJ.

We can estimate the intrinsic generation efficiency by inspecting the input and output pump spectra, as plotted in Fig. 2(b) for different conversion efficiencies. Since optical rectification is intra-pulse difference frequency generation, the IR photons can be repeatedly down converted by cascading to THz photons till the phase matching condition is violated or there is a break-up of the pump pulse, whichever happens first [7]. Consequently, the pump spectrum red shifts and broadens. The number of cascaded cycles (*N*) can be estimated based on the shift of the center of mass of the pump spectra according to:

$$N = (c/\nu_{THz}) \times (1/\lambda_1 - 1/\lambda_2) \quad (1)$$

where *c* is the speed of light in vacuum; $\nu_{THz}$ is 0.3 THz at room temperature; $\lambda_1$ is the central wavelength (center of mass) of the input spectrum at 803 nm and $\lambda_2$ is the central wavelength of the broadened IR output spectrum. For the highest pump fluence with 0.18% conversion efficiency, the output pump spectrum broadens up to 867 nm wavelength (at 20 dB lower than the peak). The center of mass of the broadened and red-shifted spectrum is located at 812 nm, which corresponds to *N*= 11 cascading cycles with an estimated intrinsic efficiency of ~0.9% ($\eta = N \cdot \hbar\nu_{THz} / \hbar\nu_{IR}$).

However, experimentally, only 0.18% conversion efficiency was extracted at room temperature. It means that many loss factors need to be considered, such as linear THz absorption, multi-photon absorption of the pump, output-coupling of the THz pulses from the crystal, etc.

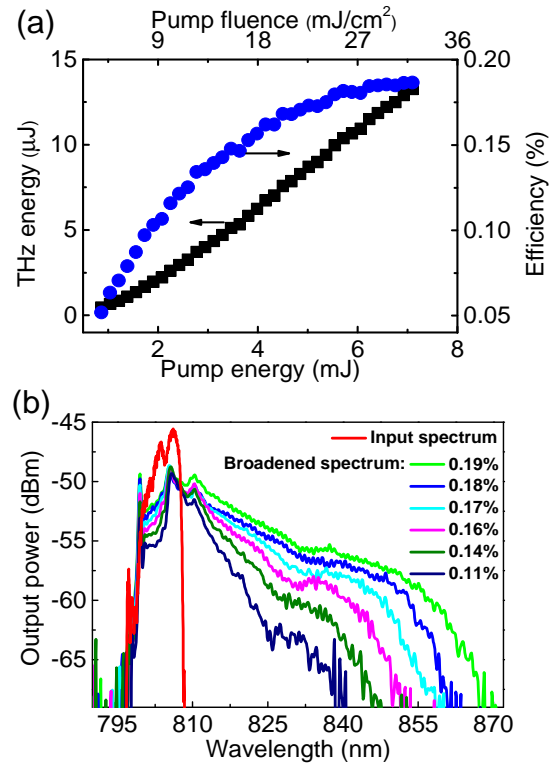

Fig. 2. (a) THz output energy and corresponding

conversion efficiency as a function of the pump energy (fluence) at room temperature. (b) Input IR spectrum and output IR broadened spectra at different conversion efficiencies.

One of the primary obstacles to achieving high conversion efficiency is the linear THz absorption [8-9]. Lithium niobate has a strong linear THz absorption due to the transverse optical (TO) phonon resonance at ~7.7 THz [10]. At room temperature, the absorption coefficient in the range of 0.1-1.0 THz is ~5.0-20 $cm^{-1}$ polarized along the optical-axis of lithium niobate [11]. Therefore, in order to reduce the THz linear absorption, cryogenic cooling of the lithium niobate crystal is crucial [12]. At a cryogenic temperature of <100 K, the absorption coefficient in this range is much lower than that at room temperature. However, the free-carrier absorption of THz radiation initiated by multi-photon absorption of the optical radiation could be important and needs more systematic investigation [13].

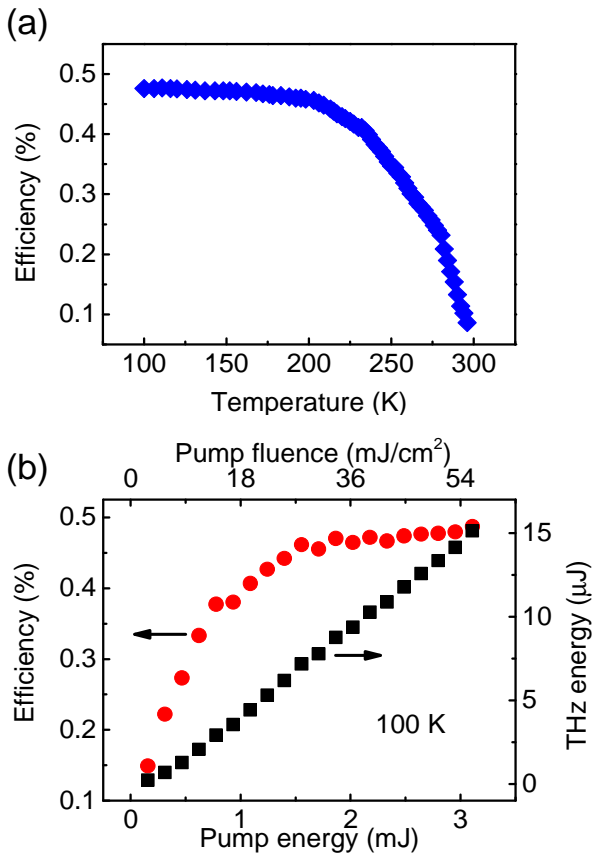

Fig. 3. (a) THz conversion efficiency as a function of crystal temperature. (b) THz energy and efficiency as a function of the pump energy at 100 K.

To obtain higher conversion efficiency, we cooled the crystal down to 100 K by using liquid nitrogen. For the cryogenic cooling experiment, we added a telescope before the grating to reduce the pump beam size and increase its fluence. As shown in Fig. 3(a), we measured the conversion efficiency as a function of the crystal temperature. We obtained a conversion efficiency of 0.5% pumped by Ti:sapphire laser systems at 800 nm. The highest conversion efficiency was obtained by optimizing the tilted pulse front setup at a peak pump fluence of 54.0 mJ/$cm^2$. The temperature dependent efficiency curve was recorded as the crystal temperature gradually heated up to room temperature. As shown in Fig. 3(a), the efficiency increases from 0.1% to 0.5% between 300 K and 100 K. The reason for the disparity in the conversion efficiency values at 300 K in Fig. 3(a) and the previously mentioned 0.18 % is because the data in Fig. 3(a) were obtained without re-optimizing the optics. This shows the importance of re-optimizing the setup for cryogenic temperatures. The refractive index of the lithium niobate crystal depends on the temperature [11] which results in a variation of the tilt-angle for phase-matching. This maximum 0.5 % value is only about half of the estimated 0.9% intrinsic efficiency. This indicates that further mitigation of THz Fresnel losses, beam transport loss, and THz absorption could result in higher conversion efficiencies. Fig. 3(b) shows the extracted THz energy and conversion efficiency at 100 K. Up to 15 µJ of THz output energy corresponding to a conversion efficiency of 0.5% was generated at a pump fluence of 54 mJ/$cm^2$ or 3 mJ of pump energy. The output THz energy varies quadratically with the pump energy or equivalently, the conversion efficiency varies linearly up to 18 mJ/$cm^2$ fluence, at which point the efficiency curve starts to saturate. We measured the imaged THz beam profile shown in Fig. 4. The 1/$e^2$ spot diameters in the horizontal and vertical directions are 1.3 mm and 1.2 mm, respectively.

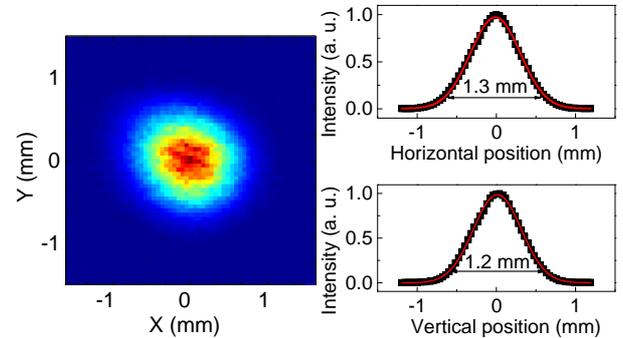

Fig. 4. THz beam profile at 100 K at constant 27 mJ/$cm^2$ pump fluence.

As mentioned before, the THz absorption coefficient for 0.3 THz is much larger at 300 K (3.3 $cm^{-1}$ at 0.4 THz) than at 100 K (1.3 $cm^{-1}$ at 0.4 THz) [8-9, 11]. Therefore, when the crystal was cryogenically cooled, the resulting THz spectrum was broader leading to shorter THz pulses. As shown in Fig. 5 (a) and (b), we measured a sub-cycle THz pulse at 100 K with a broadened THz output spectrum compared with that obtained at room temperature. The central frequency at 100 K is located at ~1.0 THz with spectral content up to 2.3 THz. At lower temperature, the THz absorption was reduced and higher frequency THz radiation is coupled out. This contributes to the higher efficiency we obtained at 100 K.

In summary, we have demonstrated a THz conversion efficiency of 0.5% with a lithium niobate crystal pumped by 150 fs Ti:sapphire laser pulses in the 800 nm wavelength range. The result was achieved by cryogenically cooling the crystal to reduce the THz absorption. Based on the calculated intrinsic efficiency, there may still be room for improving the efficiency by improving the out-coupling system, transport methods and further reducing the linear THz absorption in the crystal. Free-carrier absorption of THz radiation due to multiphoton absorption may be avoided by employing longer pump wavelengths [14]. These results are important for the rapid developing field of strong-field THz sources with promising applications in light-matter interaction physics [15] and THz based compact electron accelerators [16].

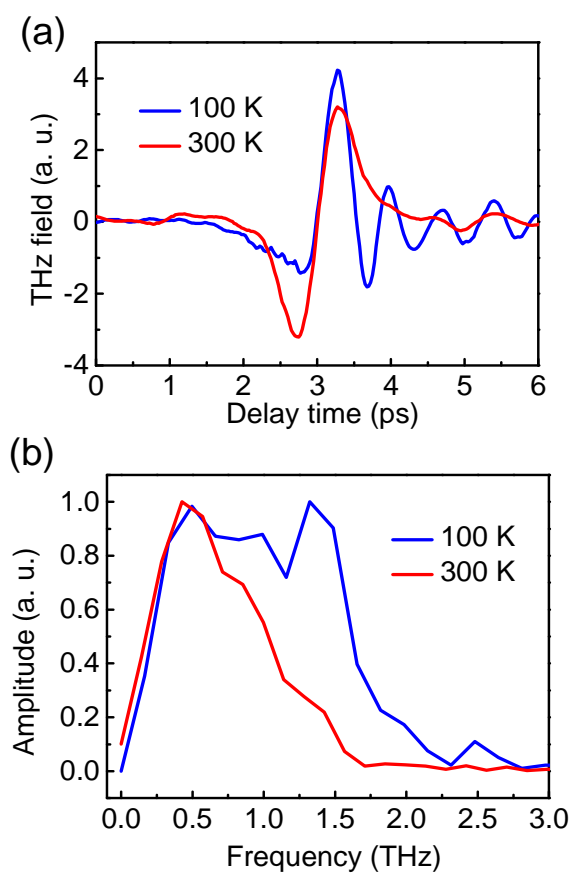

Fig. 5. (a) THz waveform measured by electro-optic sampling and (b) Corresponding THz amplitude with respect to the THz frequency.

**Funding.** This work has been supported by the excellence cluster 'The Hamburg Centre for Ultrafast Imaging - Structure, Dynamics and Control of Matter at the Atomic Scale' of the Deutsche Forschungsgemeinschaft and the Center for Free-Electron Laser Science at the Deutsches Elektronen-Synchrotron (DESY), a Center of the Helmholtz Association. We acknowledge support from the European Research Council under the European Union's Seventh Framework Programme (FP/2007-2013) / ERC Grant Agreement n. 609920. Dr. Wu acknowledges support by a Research Fellowship from the Alexander von Humboldt Foundation.